\documentclass[twocolumn,showpacs,preprintnumbers,amsmath,amssymb,aps]{revtex4-1}
\usepackage{graphicx}
\usepackage{dcolumn}
\usepackage{bm}

\begin{document}
\preprint{}
\title{Self-gravitating system made of axions}
\author{J. Barranco}%
\affiliation{Max-Planck-Institut f\"ur Gravitationsphysik 
(Albert-Einstein-Institut), Am M\"uhlenberg 1, D-14476 Golm, Germany }

\altaffiliation{Present address: Instituto de Astronom\'{\i}a, Universidad 
Nacional Autonoma de M\'exico, Mexico, DF 04510, Mexico}

\author{A. Bernal}%
\affiliation{Max-Planck-Institut f\"ur Gravitationsphysik 
(Albert-Einstein-Institut), Am M\"uhlenberg 1, D-14476 Golm, Germany}

\altaffiliation{Present address: Instituto de Ciencias Nucleares, Universidad Nacional
Aut\'onoma de M\'exico, A.P. 70-543,  04510 M\'exico D.F.,
M\'exico.}

\date{\today}
\pacs{95.35.+d,14.80.Va,04.40.-b,98.35.Gi}
\begin{abstract}
We show that the inclusion of an axion-like effective potential in the 
construction of a self-gravitating system of scalar fields  
decreases its compactness when the value of the self-interaction 
coupling constant is increased. By including the current values for the
axion mass $m$ and decay constant $f_a$, we have computed the mass and 
the radius for self-gravitating systems made of axion particles.  
It is found that such objects will have asteroid-size masses and radii 
of a few meters, thus a self-gravitating system made of axions could 
play the role of scalar  mini-MACHOs and mimic a cold dark 
matter model for the galactic halo.  
\end{abstract} 
\maketitle
The necessity of introducing dark matter (DM) as the main component of 
galactic matter has become a solid fact due to its observational support 
\cite{de Blok:2001fe}.
Nevertheless, the nature of DM is one of the most intriguing mysteries in physics. A 
large variety of particles have been considered as the main component of DM in the 
universe and only a few of them are still considered as good prospects 
since they must fulfill several requirements \cite{Taoso:2007qk}. Among the survivors, 
the neutralino and the axion are leading candidates \cite{Freedman:2003ys}.
The question we address in the present work is the following: 
If DM is mainly composed of axions, what type of astrophysical objects will the axions form?

In order to answer this question, we have solved the Einstein-Klein-Gordon (EKG) equations in 
the semiclassical limit. The source for the Einstein equations is the mean value of the 
energy-momentum tensor operator $\langle \hat T^{\mu \nu} \rangle$ of a real, quantized scalar field constructed 
with potential energy density given by \cite{Sikivie:2006ni} 
\begin{equation}\label{potential}
V(\phi)=m^2 f_a^2 \left[ 1 - \mbox{cos}\left( {\phi \over f_a} \right)\right] \, .
\end{equation}   
It is found that the resulting self-gravitating system, the axion star, will 
have an asteroid-size mass ($M\sim 10^{-16} M_{\odot}$) and radius of a few meters. 
This work improves our previous attempts \cite{Barranco:2008zzc,Barranco:2008zz} by
solving numerically the EKG system for current allowed values for the axion mass $m$ and 
decay constant $f_a$ without the necessity of any interpolation. 
Our findings differ from previous estimates where the effect of the potential 
energy density was either neglected \cite{Takasugi:1983zz} or was taken into 
account with a wrong sign in the self-interacting term of the potential 
\cite{Colpi:1986ye,Schunck:2003kk}. In the first case, it is known that there is a 
maximum mass for such self-gravitating system given by $M_{max}=0.633~ m_p^2/m$, 
where $m$ is the mass associated with the scalar field and $m_p$ is the Planck mass.
For the allowed values of the axion mass, $10^{-5}~\mbox{eV} < m < 10^{-3}~\mbox{eV}$ 
\cite{Sikivie:2006ni,Raffelt:2006cw}, the maximum mass for a self-gravitating system 
with the potential energy density (\ref{potential}) neglected 
lies in the range $10^{-8}~M\odot < M^{\mbox{axion star}}_{max} < 10^{-5}~M\odot$.  
On the other hand, when the axion is considered to have a repulsive self-interacting term,
instead of the attractive one given by (\ref{potential}),  
the maximum mass will be as big as $M \sim 10^4~M\odot$.

Here we solve the EKG system including a Taylor expansion of 
the potential energy density (\ref{potential})
and we observe that its inclusion tends to decrease the mass 
and consequently the compactness of the self-gravitating system made of axions.
Due to the smallness of the axion star's masses they could play the role 
of scalar field mini-MACHOs \cite{Hernandez:2004bm} and
they will be the final state of axion miniclusters \cite{Kolb:1993zz}
originated in the early universe at the QCD epoch \cite{Hogan:1988mp}.
Assuming that the axion is the main component of DM, 
the galactic halo will be a collisionless ensemble of axion stars and will be 
indistinguishable to the standard CDM scenario since 
$N$-body simulations of CDM with ultra-high resolution
are insensitive to particle mass granularity smaller than 
$10^5M_\odot-10^{3}M_\odot$ \cite{Ghigna:1999sn,Navarro:2008kc}.

The paper is organized as follows: in section \ref{EKG}, the EKG equations 
for a real, quantized scalar field with a Taylor expansion of the potential 
energy density (\ref{potential}) are obtained and are solved for arbitrary 
values of the axion mass $m$ and the decay constant $f_a$. In Sec. \ref{main} 
we include the current values of $m$ and $f_a$ and we obtain the mass and radius 
of the axion-stars. We finish section \ref{main} by commenting on some 
consequences derived in the case that the axion-star has the properties calculated here.
\section{Einstein-Klein-Gordon with an axion-like potential}\label{EKG}
\begin{figure}
\includegraphics[angle=270,width=0.5\textwidth]{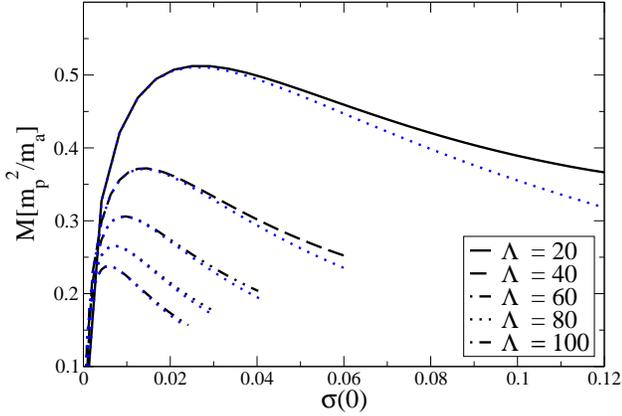}
 \caption{Gravitational mass as a function of the central value 
of the scalar field $\sigma(0)$ for different values of $\Lambda$. Dotted lines include only an expansion
of the axion potential up to the term $\Phi^4$}\label{masses}
\end{figure}
Since axions are real scalar particles, it is very useful to remember how
self-gravitating systems made of spin zero particles are constructed.
We will follow the method developed by Ruffini and Bonazzola \cite{Ruffini:1969qy}. 
The self-gravitating system arise as a solution of the EKG
equations:
\begin{equation}\label{einstein}
G_{\mu\nu}=8 \pi G < \hat T_{\mu\nu} >\,,
\end{equation}
\begin{equation}\label{KG}
\left(\Box - \frac{dV(\Phi)}{d\Phi^2}\right)\Phi=0\,,
\end{equation}
where $\Box=(1/\sqrt{-g})\partial_\mu[\sqrt{-g}g^{\mu \nu}\partial_\nu]$
and $V(\phi)$ is the scalar field potential.
Here $<\dots>$ denotes an average over the ground state of a system of many particles.
Its presence refers to the fact that we are working in the semi-classical limit
of the Einstein's equations. We will work with units where $c=\hbar=1$. 
In the case of a spherically symmetric, static space-time described by
\begin{equation}\label{metric}
ds^2=B(r)dt^2-A(r)dr^2-r^2(\sin^2\theta d\phi^2+d\theta)\,,
\end{equation}
it has been shown that such self-gravitating systems are fully 
characterized by the scalar field properties, i.e. the mass $m$ of the 
scalar field and its energy density potential $V(\Phi)$ \cite{Schunck:2003kk,Schunck:1999zu}.
The total mass of the resulting object and the typical radius depend mainly on these two 
properties of the scalar field. The axion is no exception. 
To deal with the quantum nature of the axion field, we have to
compute the average $\langle\hat T^{\mu\nu}\rangle$ in eq. (\ref{einstein}).
What is usually done is
to quantize the scalar field $\Phi \to \hat \Phi=\hat \Phi^++\hat \Phi^-$ where
\begin{eqnarray}\label{quantum}
\hat \Phi^{+}&=&\sum_{nlm} \mu_{nlm}^+R_{nl}(r)Y^l_m(\theta ,\psi)e^{-iE_nt}\nonumber\\
\hat \Phi^{-}&=&\sum_{nlm} \mu_{nlm}^-R_{nl}(r)Y^{l*}_m(\theta, \psi)e^{+iE_nt}
\end{eqnarray} 
and $\mu_{nlm}^{+(-)}$ are the usual creation (annihilation) operators for 
a particle with angular momentum $l$, azimuthal momentum $m$ and energy $E_n$.
These operators satisfy the usual commutation relations
$[\mu_{nlm}^+,\mu_{n'l'm'}^+]=[\mu_{nlm}^-,\mu_{n'l'm'}^-]=0$ and $
[\mu_{nlm}^+,\mu_{n'l'm'}^-]=\delta_{n n'}\delta_{l l'}\delta_{m m'}$.
With the operator $\hat \Phi$, it is now possible to construct the energy-momentum tensor 
operator $\hat T_{\mu \nu}$ just by inserting the operator $\hat \phi$ into the classical
expression for the energy-momentum tensor
\begin{equation}
T^\mu_\nu=g^{\mu\sigma}\partial_\sigma\phi\partial_\nu\phi-\frac{1}{2}\delta^\mu_\nu
g^{\lambda \sigma}\partial_\lambda \phi\partial_\sigma \phi-\delta^\mu_\nu V(\phi)\,.
\end{equation}
The average $\langle Q|\hat T_{\mu \nu}|Q\rangle$ is done by considering a state $|Q \rangle$ 
for which all  $N$ particles are in the ground state ($l=m=0$, $n=1$). The ground state 
satisfies $\mu^-_{100}| Q\rangle =0$. 
This procedure, as pointed out in \cite{Ruffini:1969qy}, cancels all time dependence on
the vacuum expectation value $\langle Q|\hat T_{\mu \nu}|Q\rangle$ and, for the case of a 
free scalar field ($V(\Phi)=\frac{m^2}{2}\Phi^2$), the real quantized scalar field yields
the same field equations as those obtained by using a classical complex scalar field. At this level, 
the self-gravitating system for a real quantized scalar field does not differ from a complex classical 
scalar field, hence, a real quantized scalar field does not produce the so called 
``oscillatons'' \cite{UrenaLopez:2001tw}, which are time-dependent.
In our case we are interested in the axion potential (\ref{potential}). In order
to compute $\langle \hat T^{\mu\nu}\rangle$, we perform a Taylor expansion of (\ref{potential}), 
i.e. 
\begin{equation}\label{taylor}
V(\Phi)\sim\frac{m^2}{2}\Phi^2-\frac{1}{4!}\frac{m^2}{f_a^2}\Phi^4+\frac{1}{6!}\frac{m^2}{f_a^4}\Phi^6-...
\end{equation}
We will show that the final results do not depend strongly on the number of terms considered 
in the Taylor expansion of (\ref{potential}).
The relevant term that should be considered is the
self-interacting term $\Phi^4$ and the sign it carries with itself, 
which differs from the one considered in Boson Stars (BS) \cite{Colpi:1986ye}.
With the potential (\ref{taylor}), it is possible to compute $\langle \hat T^{\mu}_\nu\rangle$ by
performing the quantization and averaging procedure previously discussed. 
The computed average of the stress energy tensor is
\begin{eqnarray}\label{vacuum}
\langle T^0_0\rangle&=&-\frac{E^2R^2}{2B}-\frac{R'^2}{2A}-\frac{m^2R^2}{2}
+\frac{m^2R^4}{12f_a^2}-\frac{m^2R^6 }{144f_a^4}+\dots \,,\nonumber \\
\langle T^1_1\rangle&=&\frac{E^2R^2}{2B}+\frac{R'^2}{2A}-\frac{m^2R^2}{2}
+\frac{m^2R^4}{12f_a^2}-\frac{m^2R^6 }{144f_a^4}+\dots\,,\nonumber \\
\langle T^2_2\rangle&=&\frac{E^2R^2}{2B}-\frac{R'^2}{2A}-\frac{m^2R^2}{2}
+\frac{m^2R^4}{12f_a^2}-\frac{m^2R^6 }{144f_a^4}+\dots\,.
\end{eqnarray}
We have dropped all sub-indexes since, as we have already pointed out, we will assume that the 
axion is in its ground state. We can observe that there is no time dependence in (\ref{vacuum}). 
Furthermore, there are new numerical factors that appear due to the averaging performed in 
$\hat T_{\mu\nu}$ \cite{Ho:2002vz}. For instance $\langle \Phi^4 \rangle = 2 R^4$ and 
$\langle \Phi^6 \rangle=5 R^6$, in such a way that we cannot recover the original $
\cos(\Phi/f_a)$ from which we departed. Following a similar procedure  but now applied 
to the scalar wave equation (\ref{KG}), with a potential (\ref{taylor}) and the 
spherically symmetric metric (\ref{metric}), the Einstein-Klein-Gordon system is obtained:
\begin{eqnarray}\label{sistema}
\frac{A'}{A^2 r}+\frac{1}{r^2}\left(1-\frac{1}{A}\right)&=&-8\pi G\langle T^0_0\rangle\,,\nonumber \\
\frac{B'}{ABr}-\frac{1}{r^2}\left(1-\frac{1}{A} \right)&=& 8\pi G\langle T^1_1\rangle \,,\nonumber \\
R''+\left(\frac{2}{r}+\frac{B'}{2B}-\frac{A'}{2A}\right)R'&+&
A\Big[\left(\frac{E^2}{B}-\frac{m^2}{2}\right)R+{} \nonumber \\  {}+
\frac{m^2R^3}{6f_a^2}-\frac{m^2R^5}{48f_a^4}\Big]&=&0\,.
\end{eqnarray}
Following standard definitions \cite{Colpi:1986ye},  we rewrite the system
(\ref{sistema}) in dimensionless variables: $x=rm$, 
$R=\sigma/\sqrt{4 \pi G}$ and $\tilde B=m^2B/E^2$,
and we have found it convenient to define the dimensionless self-interaction term
\begin{equation}\label{axionlambda}
\Lambda={1\over 24\pi}\left( {m_p \over f_a} \right)^2\,.
\end{equation}
By imposing regularity at the origin and flatness at infinity, system (\ref{sistema})
is solved using a shooting method.
Even though the set of equations (\ref{sistema}) is very similar to the case for typical BS,
\cite{Colpi:1986ye}, the behavior we found for the
family of solutions with zero-nodes is completely different.
A full set of equilibrium configurations is shown in Fig. \ref{masses}, where
the gravitational mass is plotted for different values of $\sigma(0)$ and $\Lambda$.
The equilibrium configurations have a maximum mass $M_{max}$ at some 
$\sigma(0)=\sigma_c$ for each value of $\Lambda$. But the switch in the potential sign
of the $\Phi^4$ term produces a significant change in the behavior of $M_{max}$
in comparison with standard BS \cite{Colpi:1986ye}. 
The relation  $M_{max}\sim \Lambda^{1/2}$ is no longer satisfied. Instead of 
increasing $M_{max}$ as we increase the value of $\Lambda$,
we found a decreasing $M_{max}$. 
\begin{figure}
\includegraphics[angle=270,width=0.5\textwidth]{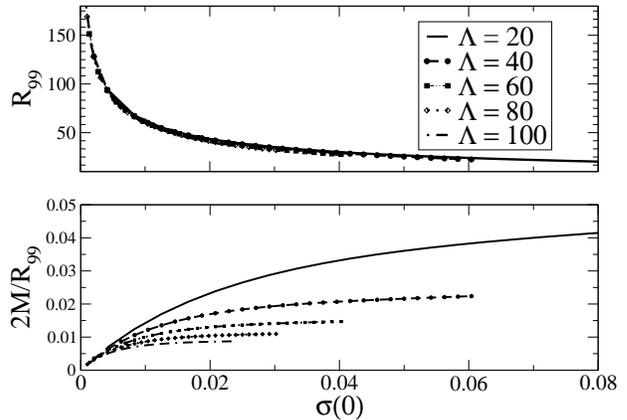}
\caption{$R_{99}$ and compactness for different configurations. The dependence of $R_{99}$ on $\Lambda$ is
negligible. For a given value of $\sigma(0)$, the compactness 
decreases as $\Lambda$ increases.}\label{compactness}
\end{figure}
This effect is expected since instead of adding a repulsive interaction between the particles 
of the system, the change in the sign due to the cosine-like potential (\ref{potential}) 
implies an attractive potential. Thus the total number of particles needed to form an 
equilibrium configuration that balance the gravitational collapse against the quantum 
pressure is lower than the case of a repulsive potential. 
One can think that this effect is apparent and as soon as the complete potential 
(\ref{potential}) is implemented in the EKG system, a different behavior would 
be seen. But the decrease in the mass of the equilibrium configurations is a robust 
behavior. The masses of equilibrium configurations including up to the fourth power of $\Phi$
in the Taylor series are plotted also in Fig. \ref{masses}, illustrating this robustness.

The bigger the value of $\Lambda$, the lower the differences on the masses. This is 
because when $\Lambda$ is increased, $\sigma_c$ decreases 
(and equivalently $\Phi(r)$ where we are interested). Then, the true expansion parameter  
of (\ref{potential}) is $\Lambda \Phi$, an it is always a small parameter .
Another interesting issue is that the dependence of the radius $R_{99}$ (defined as the radius where
$99\%$ of the total gravitational mass is reached) on the value of $\Lambda$ is weak,
as it is shown in upper panel of Fig. \ref{compactness}.
Combining the invariance of the radius as $\Lambda$ increases, with the decrease in the mass, 
means that the self-gravitating system made of a scalar field that
has an axion-like potential has a lower compactness ($2M/R_{99}$) as the self-interaction term increases.
This ``newtonization'' of the system is shown on the lower panel of Fig. \ref{compactness}.
\section{Axion-star}\label{main}
\begin{figure}
\includegraphics[angle=270,width=0.5\textwidth]{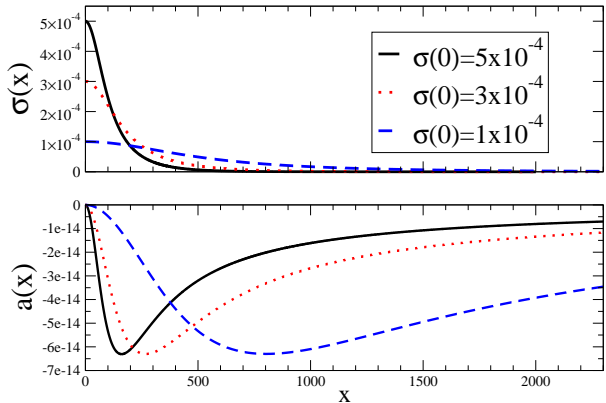}
 \caption{Scalar field and potential for a typical axion star with
different values of $\sigma(0)$ and  an axion mass of $m_a=1.0 \times 10^{-5}$eV.}\label{axionstar}
\end{figure}
The previous results where obtained assuming arbitrary values of the mass $m$
of the scalar field associated with the axion as well as free values for the decay constant $f_a$.
But the mass of the axion is constrained by astrophysical and cosmological 
considerations to lie in the range $10^{-5} ~\mbox{eV} \le m \le 10^{-3}~\mbox{eV}$ and
the decay constant is related to the axion mass by 
$m= 6 \mu \mbox{eV}\left(\frac{10^{12}\mbox{GeV}}{f_a}\right)$
\cite{Sikivie:2006ni,Amsler:2008zzb}
With these two restriction we have $10^{13}< \Lambda <10^{17}$ and then, the
previous selection of dimensionless variables $\{x,\sigma, A.\tilde B\}$) is now 
inadequate in order to numerically solve the system (\ref{sistema}). 
After some frustrated attempts, we found that a more suitable 
set of variables to solve the system (\ref{sistema}) is the
following:
\begin{equation}\label{variables}
R=\frac{f_a}{\sqrt{m}}\sigma\,, \quad r=\frac{m_p}{f_a}\sqrt{\frac{m}{4\pi}}x\,,\quad
\frac{1}{\tilde B}=\frac{E^2}{m^2 B}\,.
\end{equation}
Since $\Lambda >> 1$, it is natural to think that the resulting axion star will have
a small compactness and low mass. So, besides the change in variables (\ref{variables}),
it is convenient to solve for $a(x)=1-A(x)$. 
Solving the system (\ref{sistema}) for the new set of
variables $\{x,\sigma(x),a(x),\tilde B(x)\}$ we obtained typical 
nodeless configurations for these axion stars. Some of them are shown in Fig. \ref{axionstar}, 
where we have taken an axion mass $m=10^{-5}~$eV.
\begin{table}
\caption{Masses and $R_{99}$ for the configurations shown in Fig. \ref{axionstar}}\label{physicalvalues}
        \begin{tabular}{c|c|c|c}
            $\sigma(0)$ & Mass (Kg) & $R_{99}$ (meters) & density $\rho$ (Kg/m$^3$) \\
            \hline     
            $5 \times 10^{-4}$& $3.90 \times 10^{13}$ & $1.83$ & $6.3 \times 10^{12}$\\
            $3 \times 10^{-4}$  &$6.48\times 10^{13}$ & $2.86 $ &$2.7 \times 10^{12}$ \\
            $1 \times 10^{-4}$ &$1.94\times 10^{14}$ & $8.54 $ & $3.1 \times 10^{11}$ \\
            \hline
         \end{tabular}  
\end{table}
The total gravitational mass and the radius $R_{99}$, both in physical units, for those
configurations are shown in Table \ref{physicalvalues}.

A possible scenario emerges with the hypothesis that DM is mainly composed by axions.
As was already pointed out by Kolb and Tkachev \cite{Kolb:1993zz}, nonlinear  effects in 
the evolution of the axion field in the early universe may lead to the formation of 
``axion miniclusters''. These miniclusters may relax, due to the collisional $2a \to 2a$ 
process or by gravitational cooling \cite{Seidel:1993zk}, and they will evolve to a BS. 
In the present work we have constructed those BSs for axion particles by 
solving the EKG system for a real quantized scalar field which is regulated by an axion-potential 
(\ref{potential}). These self-gravitating systems, the axion stars, have very small masses and 
radii of meters (Table \ref{physicalvalues}) and consequently very low compactnesses. 
The resulting densities are not enough to produce stimulated decays of the axion to 
photons since they occur when $\Gamma_\pi m_p^2 V_e f_\pi /(R m_\pi^4 f_a) > 1$ which 
implies densities $\rho > 10^{15}$ Kg/ m$^3$ for $m=10^{-5}~$eV \cite{Tkachev:1987cd,Seidel:1993zk}. 
Typical densities for axion stars are shown in table \ref{physicalvalues}.

The galactic halo will be an ensemble of axion stars and this picture is not in contradiction 
with observations since the size of axions stars fit into the limits coming from microlensing  
or gravothermal instability \cite{Hernandez:2004bm}. Previous studies that construct models
for galactic dark matter halos out of scalar fields assume that each galactic halo is a spherical
Bose-Einstein condensate made of an ultra-light scalar field 
\cite{SJSin,LeeKoh,MatosEtAl2000,MatosGuzman2001,Arbey,bosonS}. In the present work, 
the axion stars will play the role of the scalar field mini-MACHOs, that is, a scenario
where the scalar field (the axion) form a large number of stable asteroid-sized 
scalar condensations which end up clustering into structures similar to CDM halos with all
their advantages. The stability of structures made of scalar fields have been extensively studied  
\cite{SeidelSuen90,HawleyChoptuik00,Guzman:2004jw}, and those analysis could be extended to 
our case with potential energy density given by (\ref{potential}).

Furthermore, axion stars present similar characteristics to the recently proposed neutralino stars \cite{Ren:2006tr}, 
with the advantage that they could survive longer periods of time \cite{Dai:2009ik}.

If DM is distributed  as axion stars, their detection will be very difficult. The proposed 
femptolensing to detect axion compact objects \cite{Kolb:1995bu} is close to its lower 
detectable limit. Another related issue is the low number of axion stars around the earth. 
Assuming for instance a Navarro-Frenk-White profile for the galactic halo, and a local 
halo density of $0.3 ~\mbox{GeV}/$cm$^3$ around the Sun, there will be $\sim 1$ axion 
star in the volume cover between Jupiter and the Sun.
Nevertheless, another axion properties can shed light on the axion, such as the conversion 
of axions into photons in the presence of strong magnetic fields \cite{Lai:2009hp}. 
Collisions of axion stars with neutron stars \cite{Iwazaki:2000hq} will produce 
flashes of light that could be detected by Gamma ray Observatories \cite{Sreekumar:1997un}.
A more detailed analysis of these ideas together with a more detailed study                 
of the stability of the axion stars could help us to determine if  DM is mainly 
composed by scalar field particles as the axions.\\

{\bf Acknowledgments: }
We would like to thank Tonatiuh Matos, Luis Urena, Shin Yoshida and 
Carlos Palenzuela for very enlightening discussion and Aaryn Tonita for 
careful reading of the manuscript. We thank L. Rezzolla for his support and hospitality at the AEI.
This work was supported in part by the CONACYT and CONACYT-SNI.

%
%

\end{document}